\documentclass[usenatbib,useAMS]{mn2e}
\usepackage{times}
\usepackage{epsfig}
\usepackage{amsmath}
\usepackage[usenames,dvips]{color}
\graphicspath{{fig/}}

\begin{document}
\title[Orbital motion effects in astrometric microlensing]{Orbital motion effects in astrometric microlensing}
\author[Sedighe Sajadian]
{Sedighe Sajadian \thanks{sajadian@ipm.ir} \\
School of Astronomy, IPM (Institute for Research in Fundamental
Sciences), P.O. Box 19395-5531, Tehran, Iran}

\maketitle
\begin{abstract}
We investigate lens orbital motion in astrometric microlensing and
its detectability. In microlensing events, the light centroid shift
in the source trajectory (the astrometric trajectory) falls off much
more slowly than the light amplification as the source distance from
the lens position increases. As a result, perturbations developed
with time such as lens orbital motion can make considerable
deviations in astrometric trajectories. The rotation of the source
trajectory due to lens orbital motion produces a more detectable
astrometric deviation because the astrometric cross-section is much
larger than the photometric one. Among binary microlensing events
with detectable astrometric trajectories, those with stellar-mass
black holes have most likely detectable astrometric signatures of
orbital motion. Detecting lens orbital motion in their astrometric
trajectories helps to discover further secondary components around
the primary even without any photometric binarity signature as well
as resolve close/wide degeneracy. For these binary microlensing
events, we evaluate the efficiency of detecting orbital motion in
astrometric trajectories and photometric light curves by performing
Monte Carlo simulation. We conclude that astrometric efficiency is
$87.3$ per cent whereas the photometric efficiency is $48.2$ per
cent.


\end{abstract}
\section{Introduction}
Gravitational field of an object acts as a gravitational lens which
deviates light path of a collinear background source and produces
distorted images \cite{Einstein36}. In Galactic scale, the angular
separation of the images is of order of milli-acrsecond that is too
small to be resolved even by the most modern telescopes. Instead,
the combined light of images received by an observer is magnified in
comparison to the un-lensed source. This phenomenon is called
gravitational microlensing which was proposed as a method to probe
dark objects in Galactic halo, extra solar planets, study stellar
atmosphere, etc.\cite{Leibes,ChangRefesdal,Paczynski86}.

One of features of gravitational microlensing is the deviation of
the light centroid of the source star images from the source star
position. For the case of a point-mass lens, the centroid shift of
source star images traces an ellipse in the lens plane while the
source star is passing an straightforward line in the lens plane
\cite{Walker,Miyamoto,Hog,Jeong}. In contrast with the magnification
factor which is a dimensionless scalar, the light centroid shift of
the source star images is a dimensional vector and its size is
proportional to angular Einstein radius, i.e. the angular radius of
images ring when observer, source and lens are completely aligned.
For a stellar-mass lens, the angular Einstein radius is a few
hundreds micro-arcsecond, too small to be observed directly.
However, its quantity enhances in two cases: (i) when the lens mass
is high, e.g. a stellar-mass black hole as microlens and (ii) when
the lens is very close to the observer. The light centroid shift of
source star images in these microlensing events is measurable by
high-precision interferometers e.g. Very Large Telescope
Interferometry (VLTI) or the HST \cite{Paczynski95}. An important
future project is GAIA, performing high precision astrometry, which
will be operational in near future \cite{GAIA}.

During a microlensing event by measuring the astrometric lensing and
parallax effects, the mass of the deflector can be inferred without
knowing the exact distances of the lens and the source from the
observer \cite{Paczynski97,Miralda96}. Also the degeneracy in
close/wide caustic-crossing binary microlensing events
\cite{Dominik99} can be removed by astrometric measurements
\cite{Gould,Chung}. The two configurations of the lens and source
produce different astrometric trajectories even though they might
present the same light curves. Hence, by measuring the astrometric
trajectories this degeneracy can be removed \cite{Hanetal99}.

Generally, there are some anomalies in microlensing events which
deviate the observed light curve and astrometric trajectory of
source star from the standard model. Some deviations owing to these
anomalies help obtain extra information about lens and source and
thus resolve degeneracy. One of them is the effect of lens orbital
motion in a binary microlensing event. In this case two components
in a gravitationally bound binary system which act as lens rotate
around their common center of mass. As a result, their orientation
changes as a microlensing event progress and the resulting light
curve is different from that due to a static binary microlens. The
ratio of the Einstein crossing time to the orbital period of lens
system is considered as a criterion which indicates the probability
of detecting the orbital motion in the microlensing light curve
\cite{Dominik98,Ioka99}. Measuring this effect gives information
about the orbit of microlenses system which in turn helps to resolve
the close/wide degeneracy \cite{An2002,Gaudi2008,Dong2009,Shin2013}.
Detectability of orbital motion in light curves of binary
microlensing events was investigated extensively by Penny et al.
(2011) whereas the relevant deviation in the astrometric trajectory
of source star and its detectability are not clear yet.

In this work, we study lens orbital motion in astrometric
microlensing and its detectability. In microlensing events, the
astrometric trajectory of source star tends to zero much more slowly
than the light amplification as the source distance from the lens
position rises. As a result, perturbations enlarged with time such
as lens orbital motion can produce considerable deviations in the
astrometric trajectories. Since the astrometric cross-section is
much larger than the photometric one, so the rotation in the source
trajectory owing to the orbital motion is more probable to be
detected in the astrometric trajectory than the photometric light
curve. However, depending on the size of the angular Einstein radius
this effect can be detected. We find that, among binary microlensing
events with detectable astrometric trajectories, those with rotating
stellar-mass black holes have most likely detectable astrometric
signatures of orbital motion. For these binary microlensing events,
we evaluate the efficiency of detecting orbital motion in
astrometric trajectories and photometric light curves by performing
Monte Carlo simulation. Detecting lens orbital motion in their
astrometric trajectories helps to discover even secondary components
with no photometric binarity signatures as well as resolve
close/wide degeneracy.

This paper is organized as follows: In section (\ref{one}) the
effect of orbital motion in binary microlensing events are
explained. We next describe astrometric properties of microlensing
events and finally we investigate the orbital motion effect on the
astrometric trajectories. In the next section by preforming a Monte
Carlo simulation we study the astrometric and photometric
efficiencies for detecting lens orbital motion. In section
(\ref{result}) we explain the results.

\section{Orbital motion in astrometric microlensing}
\label{one} In this section we first study the orbital motion effect
in binary microlensing events. Having reviewed the astrometric
microlensing, we investigate the microlensing events with detectable
astrometric shifts in source trajectories. Finally, we study whether
the orbital motion effect in the astrometric trajectories is
detectable.

\subsection{Orbital motion effect of binary microlenes}
Dominik (1998) first studied the effect of lens orbital motion in
binary microlensing events and concluded that in the most
microlensing events this effect is ignorable, but in some
long-duration microlensing events can probably be observed. Orbital
motion effects on the planetary signals in high-magnification
microlensing events was investigated by Rattenbury et al. (2002).
Recently, Penny et al. (2012) indicated how fraction of binary and
planetary microlensing events exhibit the orbital motion effects in
their light curves by performing Monte Carlo simulation and
investigated those factors which change this fraction. Until now,
several observed microlensing events have shown the signatures of
lens orbital motion. In some cases, these signatures have allowed to
infer some of orbital parameters and remove the close/wide
degeneracy \cite{An2002,Gaudi2008,Dong2009,Shin2013}. Sometimes,
there are several degenerate best-fitting solutions for orbital
motion modeling. Three systems with complete orbital solutions have
been found, e.g. the microlensing event OGLE-2011-BLG-0417
\cite{Shin11,Shin12}. However, one of these solutions can actually
be checked with radial velocity method \cite{Gould13}. In the
following we study the orbital motion effect in microlensing light
curves in the same way as Dominik's approach.

Let us consider a gravitationally bound system as microlens. Under
the gravitational effect, the secondary component with respect to
the first one traces out an ellipse one of whose focus is at the
center of mass position. Two components of their relative distance
are given by:
\begin{eqnarray}
x(\xi) &=& a(\cos \xi -\varepsilon) \nonumber\\
y(\xi) &=&a\sqrt{1-\varepsilon^2}\sin \xi,
\end{eqnarray}
where $x$ and $y$ axes are along the semi-major and semi-minor axes,
$\varepsilon$ is the eccentricity, $a$ is the semi-major axis and
$\xi$ is a periodic function with time which for a small
eccentricity and up to its first order is estimated as:
\begin{eqnarray}
\xi(t)=2 \pi\frac{t-t_{p}}{P}+2 \varepsilon \sin(2 \pi
\frac{t-t_{p}}{P}),
\end{eqnarray}
where $t_{p}$ is the time of arriving at the perihelion point of
orbit, $P$ is the orbital period of the lenses motion. To study the
orbital motion of lenses on microlensing light curves, we should
project the orbit plane of lenses into the sky plane. In that case,
we need to two consecutive rotation angles: $\beta$ around $x$-axis
and $\gamma$ around $y$-axis. The projected components of the
relative position vector of the second one with respect to the
first, in the sky plane normalized to the Einstein radius are:
\begin{eqnarray}\label{eqx}
x_{1}(t)&=&\rho[\cos\gamma(\cos\xi(t)-\varepsilon)+ \nonumber\\
&&+\sin\beta\sin\gamma\sqrt{1-\varepsilon^2}\sin\xi(t)] \nonumber\\
x_{2}(t)&=&\rho\cos\beta\sqrt{1-\varepsilon^2}\sin\xi(t),
\end{eqnarray}
where $\rho$ is the normalized semi-major axis to the Einstein
radius \cite{Dominik98}. Hence, lens orbital motion causes the
projected distance between two lenses in the lens plane changes with
time which is given by:
\begin{eqnarray}
d(t)=\sqrt{x_{1}(t)^2+x_{2}(t)^2}.
\end{eqnarray}
On the other hand, the binary axis with respect to the projected
source trajectory rotates with angle
$\theta_{l}(t)=\tan^{-1}(x_{2}/x_{1})$. So, during a microlensing
event with binary rotating lenses, the source trajectory rotates
with respect to the binary axis with angle $-\theta_{l}$ around the
line of sight towards the observer. However, this rotation is not
always detectable in the observer reference frame.

\begin{figure}
\begin{center}
\psfig{file=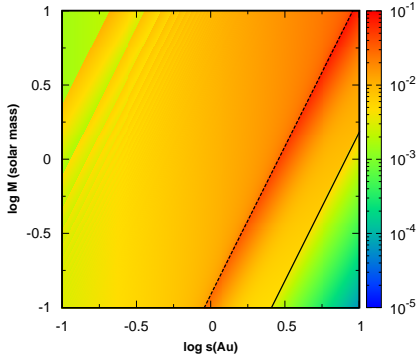,angle=270,width=8.cm,clip=} \caption{The map
of the detectability factor $f$ in the lenses parameter space
containing the lenses total mass $M$ and the semi-major axis of
lenses orbit $s$. The black dashed and solid lines represent the
branch values of the projected distance between two lenses
normalized to the Einstein radius between $3$ different topologies
of caustic: close, intermediate and wide binaries respectively.}
\label{fig1}
\end{center}
\end{figure}

The signature of lens orbital motion in the microlensing light
curves is not always observable and its Detectability depends
strongly on the ratio of the lensing characteristic time to the
orbital period of lenses motion. Dominik (1998) considered the
Einstein crossing time as the characteristic time of a binary
microlensing event to evaluate the orbital motion detectability.
However, the orbital motion detection efficiency for binary
microlensing events without caustic-crossing features is too small
and of order of one per cent \cite{penny11}. As a result, we
consider the effective time scale for detecting signatures of lens
orbital motion as $\tilde{t_{E}}= \Delta(q,d)t_{E}$ instead of the
Einstein crossing time, where $\Delta(q,d)$ refers to the average
over the ratios of the lengths of the source trajectories providing
the sources are inside the caustic curve to the total length of the
source trajectories. We assume that there are several parallel
source trajectories over the lens plane with the same lengths i.e.
$L$ so that they cover the caustic curve. In this case,
$\Delta(q,d)$ is given by:
\begin{eqnarray}\label{ddq}
\Delta(q,d)=\frac{1}{L}(\sum^{N}_{i}l_{i}/N)=\frac{1}{N L
\delta}\sum^{N}_{i}l_{i}\delta= \mathcal{S}_{c},
\end{eqnarray}
where $l_{i}$ is the length of the portion of $i$th source
trajectory providing the source is inside the caustic curve, $N$ is
the number of the source trajectories, $\delta=L/N$ is the normal
distance between two consecutive source trajectories, $d$ is the
projected distance between two lenses normalized to the Einstein
radius and $q$ is the ratio of the lens masses. According to
equation (\ref{ddq}), it is clear that $\Delta(q,d)$ is equal to the
ratio of the area interior to the caustic curve to an area equal to
$L^{2}$ i.e. $\mathcal{S}_{c}$. By considering this factor, the
detectability of lens orbital motion can be characterized by the
factor $f$:
\begin{eqnarray}
f=\frac{ \tilde{t_{E}}}{P}.
\end{eqnarray}
The map of detectability factor $f$ is shown in Figure (\ref{fig1})
in the parameter space of lenses containing the total mass of lenses
$M$ and semi-major axis $s$.
In this figure the black dashed and solid lines represent the
threshold values between the close-intermediate (or resonance) and
intermediate-wide binaries respectively. For calculating the inner
area of caustics, we first determine the caustic points which have
so small determinant of Jacobian matrix. Then, after sorting the
consecutive points on the caustic line we calculate the interior
area of caustics numerically. The detectability factor has the
highest amount for the intermediate microlensing events which locate
near to the close-intermediate threshold line (dashed line).
However, Gaudi (2012) intuitively pointed out that the resonance
microlensing events according to their large sizes and
cross-sections are more sensitive to the small changes in $d$ owing
to lens orbital motion. According to this figure, the orbital motion
detectability in microlensing events with more massive lenses which
have the same caustic curve, is larger than those with less massive
lenses. For this plot, we consider $q=1$, $v_{t}=175 km/s$, $D_{l}=4
Kpc$ and $D_{s}=8 Kpc$. However, the smaller amounts of $q$ just
move two threshold lines towards each other so that they mostly form
close and wide binaries while the detectability factor is maximum
for intermediate binaries similar to the plotted case.

In the following, we explain the astrometric properties of
microlensing and seek binary microlensing events which have
detectable astrometric trajectories as well as very likely
detectable orbital motion signatures.
\begin{figure}
\begin{center}
\psfig{file=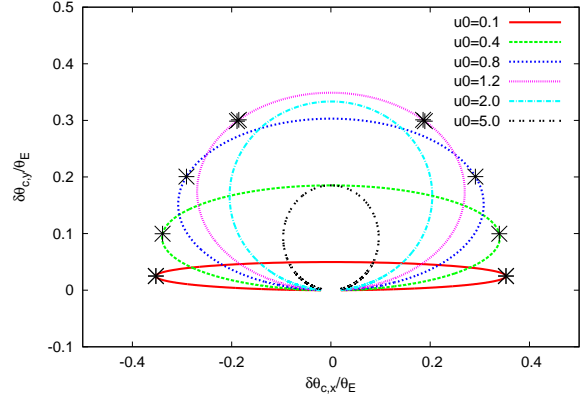,angle=270,width=8.cm,clip=} \caption{The
trajectory of the light centroid shift normalized to the
$\theta_{E}$ assuming a point-like source star is passing an
straightforward line in the gravitational field of a point-mass lens
for various amounts of impact parameter. The star signs indicate the
most deviation in the light centroid of images with respect to the
source position which happens at $u=\sqrt{2}$.} \label{figas}
\end{center}
\end{figure}

\subsection{Astrometric properties of microlensing}
In microlensing events, the light centroid vector of source star
images does not coincide with the source position. This astrometric
shift in source star position changes with time as a microlensing
event progresses. For a point-mass lens, the centroid shift vector
of source star images is given by:
\begin{eqnarray}\label{astro1}
\boldsymbol{\delta\theta}_{c}=\frac{\mu_{1}\boldsymbol{\theta_{1}}+
\mu_{2}\boldsymbol{\theta_{2}}}{\mu_{1}+\mu_{2}}-\boldsymbol{u}\theta_{E}
=\frac{\theta_{E}}{u^{2}+2}\boldsymbol{u},
\end{eqnarray}
where $\boldsymbol{\theta}_{i}$ and $\mu_{i}$ are the position and
magnification factor of $i$th image,
$\boldsymbol{u}=p\hat{x}+u_{0}\hat{y}$ is the vector of the
projected angular position of the source star with respect to the
lens normalized by the angular Einstein radius of the lens
$\theta_{E}$ in which $p=(t-t_{0})/t_{E}$ , $t_{0}$ is the time of
the closest approach, $t_{E}$ is the Einstein crossing time,
$\hat{x}$ and $\hat{y}$ are the unit vectors in the directions
parallel with and normal to the direction of the lens-source
transverse motion. The angular Einstein radius of lens is given by:
\begin{eqnarray}\label{Eins}
\theta_{E}=\sqrt{\kappa ~M_{l} ~\pi_{rel}} = 300 \mu
as\sqrt{\frac{M_{l}}{0.3 M_{\odot}}}
\sqrt{\frac{\pi_{rel}(mas)}{0.036}},
\end{eqnarray}
where $M_{l}$ is the lens mass, $\kappa=\frac{4G}{c^{2} Au}$ and
$\pi_{rel}=1Au(\frac{1}{D_{l}}-\frac{1}{D_{s}})$ where $D_{l}$ and
$D_{s}$ are the lens and source distances from the observer.
Usually, a microlensing parallax is defined as $\pi_{E}=\pi_{rel}/
\theta_{E}$ which can be measured from the photometric event.
\begin{table*}
\begin{center}
\begin{tabular}{|c|c|c|c|c|c|c|c|c|c|c|c|c|}
&  & $M(M_{\odot})$  & $D_{l}(Kpc)$ & $D_{s}(Kpc)$ & $R_{E}(Au)$ & $\theta_{E}(milli as)$ & $\mu_{l}(''/yr)$ & $V_{t}(Km/s)$ & $t_{E}(day)$ & $s(Au)$ & $P(day)$ & \\
\hline\hline

&$(\mathcal{A})$& $\geq5$ & $\sim6.5$ & $\sim8.0$ & $\geq7.0$ & $\geq1.1$ & $\sim0.006$ & $\sim175$ & $\geq70$ & $\sim4$ & $\leq1300$ & \\
\hline

&$(\mathcal{B})$& $\sim0.3$ & $\leq0.1$ & $\sim8.0$ & $\leq0.5$ & $\geq4.9$ & $\geq0.4$ & $\sim185$ & $\leq5$ & $\sim4$ & $\sim5000$ & \\
\end{tabular}
\end{center}
\caption{The possible ranges of physical parameters of two sets of
binary microlensing events with measurable shifts in the astrometric
light centroid of source star images: $(\mathcal{A})$ gravitational
microlensing events with binary stellar-mass black holes and
$(\mathcal{B})$ those with high proper motion stars located in
distances smaller than $100$ parsec from the sun.}\label{tab1}
\end{table*}

The components of this centroid shift vector are given by:
\begin{eqnarray}
\delta\theta_{c,x}(u_{0},p)&=&\frac{p}{u_{0}^{2}+p^{2}+2} \theta_{E},\nonumber\\
\delta\theta_{c,y}(u_{0},p)&=&\frac{u_{0}}{u_{0}^{2}+p^{2}+2}
\theta_{E}.
\end{eqnarray}
The shift in the light centroid trajectory of the source star images
owing to a point-mass lens traces an ellipse while the source star
is passing an straightforward line in the lens plane, plotted in
Figure (\ref{figas}). By defining $X=\delta \theta_{c,x}$ and
$Y=\delta \theta_{c,y}-\frac{u_{0} \theta_{E}}{2(u_{0}^{2}+2)}$,
these coordinates satisfy the following equation:
\begin{eqnarray}
X^{2}+(\frac{Y}{b})^2=a^{2},
\end{eqnarray}
where $b=\frac{u_{0}}{\sqrt{u_{0}^{2}+2}}$ and
$a=\frac{\theta_{E}}{2\sqrt{u_{0}^{2}+2}}$. This ellipse is called
the \emph{astrometric ellipse} \cite{Walker,Jeong}. The ratio of the
axes of this ellipse is a function of impact parameter, so that for
the large impact parameter this ellipse converts to a circle whose
radius decreases by increasing impact parameter and for small
amounts, it becomes a straight line (see Figure \ref{figas})
\cite{Walker}. since $\frac{a}{b}=\frac{2u_{0}}{\theta_{E}}$ and we
can simply obtain $u_{0}$ from photometry, so from astrometric data
we get the angular Einstein radius from which as well as the
parallax measurement the lens mass can be inferred. In this figure
the points with the maximum amounts of the centroid shift (i.e.
$u=\sqrt{2}$) were shown with star symbols. For $u\gg \sqrt{2}$ the
size of the astrometric centroid shift falls off as $\theta_{E}/u$
whereas the magnification factor of a point-like source star being
lensed by a point-mass lens for $u\gg 1$ tends to $1+2/u^{4}$.
Hence, the centroid shift of source star images falls off much more
slowly than the light amplification of source while the source
distance from the lens rises \cite{DominikSahu}. As a result, the
astrometric cross-section is larger than the photometric one in
microlensing events.

In contrast with the magnification factor which is a dimensionless
scalar, the light centroid shift of the source star images is a
dimensional vector and its size is proportional to the angular
Einstein radius given by equation (\ref{Eins}). The angular Einstein
radius of a K- or M-dwarf star as the most probable lens with
$M_{l}\sim 0.3 M_{\odot}$ located in the Galactic disk $D_{l}\sim
6.5$Kpc, in the observations towards the Galactic bulge i.e.
$D_{s}\sim 8.0$Kpc, is of order of a few hundred micro-arcsecond
(see equation \ref{Eins}) which is too small to be observed.
However, for two sets of microlensing events the angular Einstein
radius enhances: $(\mathcal{A})$ when the lens mass is high and
$(\mathcal{B})$ when the lens is so close to the observer. These two
classes of microlensing events with detectable astrometric
trajectories have different properties owing to different amounts of
the Einstein radius. In table (\ref{tab1}) we briefly express the
possible ranges of amounts of their physical parameters.

The set $(\mathcal{A})$ of microlensing events which are
characterized in the second row of table (\ref{tab1}) are
long-duration microlensing events with no light from the lens. These
events with detectable astrometric trajectories are best candidates
to indicate the mass of stellar-mass black holes located in the
Galactic disk through the photometric and astrometric data as well
as the parallax effect \cite{Sahu}. It has been five years since the
HST started observing some of long-duration microlensing events with
no light from the lens. Because no isolated black hole has been
discovered to date, so these events very likely show binarity
signatures \cite{Sahu2}. The binary microlensing events due to
stellar-mass black holes located in Galactic disk with the total
mass of lenses larger than $5 M_{\odot}$ and semi major axis of
order of $4$Au which have the orbital period larger than $
1300$days, are characterized by the Einstein radius larger than
$7.8$Au and the Einstein crossing time larger than $80$days.
Therefore, most of these events are close and a few of them are
intermediate or wide binary microlensing events in which the orbital
motion effect of lenses is not ignorable due to the considerable
ratio of the Einstein crossing time to the orbital period and the
finite size effect is too small owing to the large Einstein radius.
\begin{figure}
\begin{center}
\psfig{file=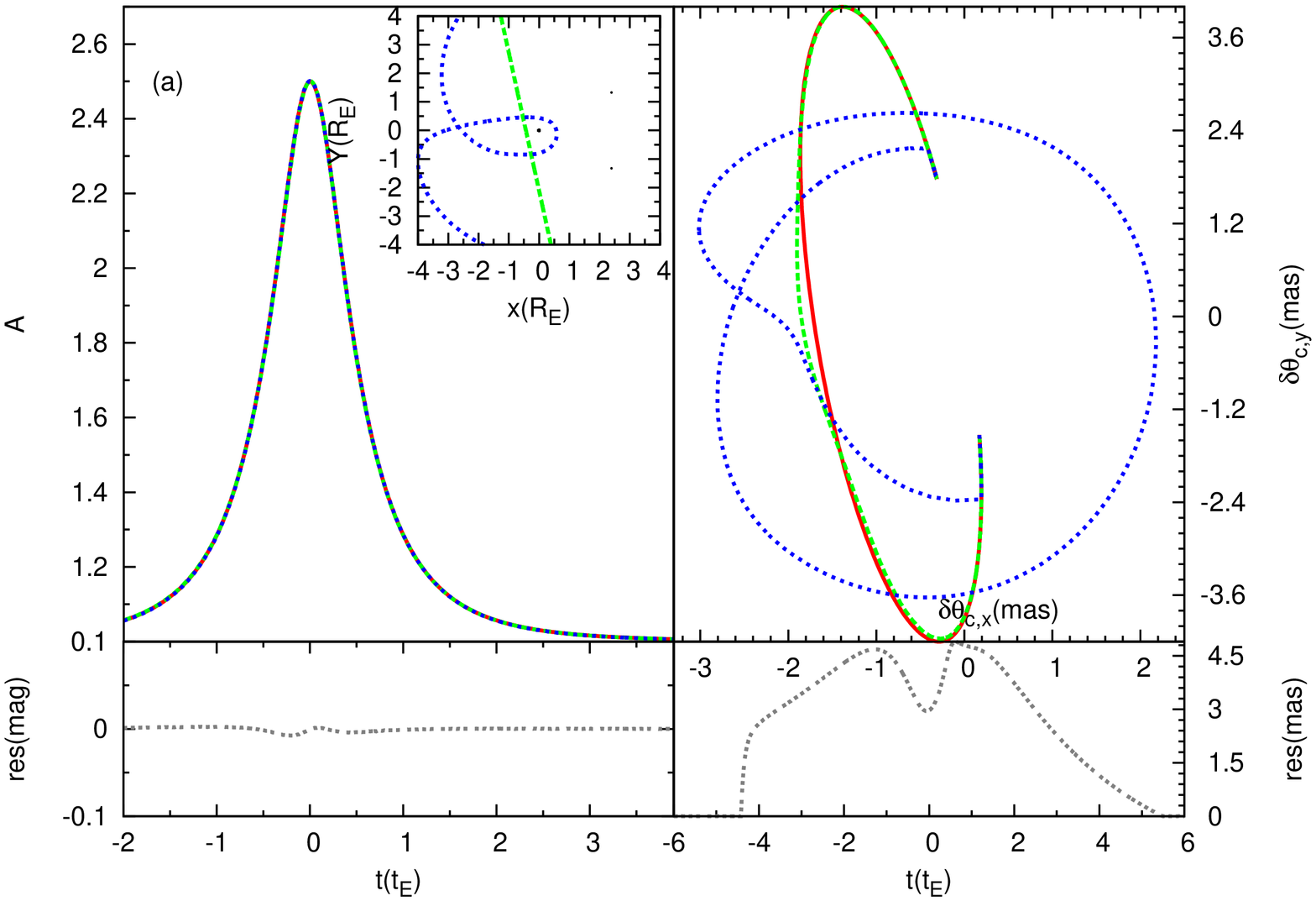,angle=270,width=8.cm,clip=}
\psfig{file=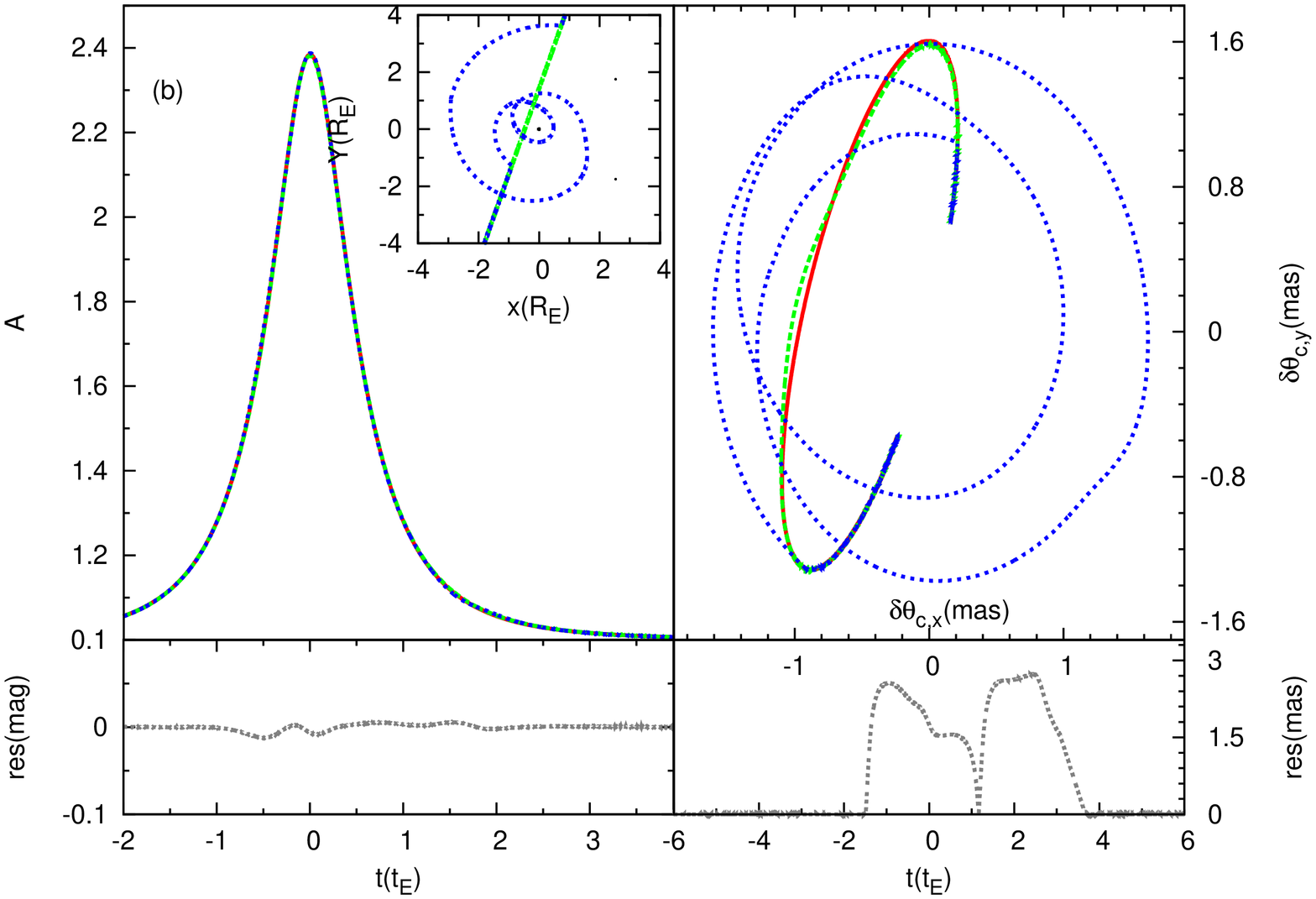,angle=270,width=8.cm,clip=}\caption{Two
typical binary microlensing events with rotating stellar-mass black
holes. In each subfigure, the light curves (left panels),
astrometric trajectories (right panels) and the source trajectories
with respect to the caustic curve (insets in the left-hand panels)
without and with considering the effect of lens orbital motion are
shown with green dashed and blue dotted lines respectively. The
point-mass lens models are shown by red solid lines. The photometric
and astrometric residuals with respect to the static binary model
are plotted with gray dotted lines. The relevant parameters can be
found in Table (\ref{tab2}).} \label{fig2}
\end{center}
\end{figure}

\begin{table*}
\begin{center}
\begin{tabular}{|c|c|c|c|c|c|c|c|c|c|c|c|c|c|}
&  & $M_{l}(M_{\odot})$ & $q$ & $s(Au)$ & $d(R_{E})$ & $t_{E}(day)$ & $P(day)$ & $\beta^{\circ}$ & $\gamma^{\circ}$& $\alpha^{\circ}$ & $u_{0}$ & $v_{t}(km/s)$ & \\
\hline\hline
& $(a)$ & $8.5$ & $0.06$ & $2.5$  & $0.33$ & $48.6$ & $478.9$ & $-5.3$ & $85.7$ & $101.7$ & $0.43$ & $271.6$ & \\
\hline
& $(b)$ & $13.7$ & $0.09$ & $3.9$ & $0.28$ & $276.7$ & $734.8$ & $-23.1$ & $69.3$ & $71.6$ & $0.45$ & $88.6$ & \\
\hline
& $(c)$ & $15.4$ & $0.12$ & $9.5$  & $0.64$ & $233.1$ & $2572.1$ & $40.3$ & $-82.7$ & $233.4$ & $0.77$ & $110.0$ & \\
\hline
\end{tabular}
\end{center}
\caption{The parameters used to make the microlensing light curves
shown in Figure (\ref{fig2}) and Figure (\ref{fig4}).}\label{tab2}
\end{table*}

The second set $(\mathcal{B})$ which are characterized in the third
row of table (\ref{tab1}) contains gravitational microlensing events
in which the lens is so close to the sun, e.g. closer than $100$
parsec from the sun position. These stars are best candidates for
the astrometric microlensing through GAIA mission for accurately
indicating their mass\cite{GAIA}. Owing to the small distance of the
lens from the sun, their proper motion is so high and larger than
$0.5$ arcsecond in year. These microlensing events are characterized
by the Einstein radius smaller than $0.5$Au and Einstein crossing
time smaller than $5$days. Let us assume that the lens is a binary
system. In that case, most of these binary microlensing events are
wide and a few of them are intermediate or close microlensing events
in which the orbital motion effect of lenses is ignorable due to the
too small ratio of the Einstein crossing time to the orbital period
and the finite-lens effect is probably considerable owing to the
small Einstein radius. Note that, the finite-source effect is mostly
ignorable because of the small amount of the relative distance of
the lens to the source star from the observer i.e.
$x(=D_{l}/D_{s})\leq 0.0125$.

Hence, binary microlensing events with rotating stellar-mass black
holes located in the Galactic disk which have detectable astrometric
shifts in the source star trajectories have most likely detectable
orbital motion signatures (case $\mathcal{A}$). In these events, the
ratio of the Einstein crossing time to the orbital period is high.
If there is no caustic-crossing feature in their light curves, the
photometric signature of orbital motion or even of the secondary
lens is too small to be detected whereas the astrometric signature
owing to the orbital motion can be detected in the astrometric
trajectories which can show the existence of a secondary component.
In this work we pay attention to these events owing to much
probability of detecting astrometric signatures of orbital motion.

\subsection{Astrometric microlensing with rotating stellar-mass black holes}
Here, we study the astrometric microlensing due to rotating
stellar-mass black holes. As mentioned, lens orbital motion changes
the lenses orientation. As a result, the magnification pattern
changes with time owing to the time variation of the projected
distance between two lenses as well as the straightforward source
trajectory rotates with respect to the binary axis. However, source
star does not always receive the gravitational effect of the
secondary component, e.g. when the ratio of the lens masses is so
small. In that case, the source trajectory does not change in the
observer reference frame. Considering this point, we let the lenses
rotate around their common center of mass when the astrometric
trajectory deviates more than ~5 micro arcseconds from the
astrometric ellipse due to a point-mass lens located at the center
of mass position whose mass is equal to the total mass of the
lenses. However, when this difference becomes smaller than the
threshold amount, we stop the lenses rotation and let the source
trajectory slowly return back to its straightforward trace.

In Figure (\ref{fig2}) we represent two examples of microlensing
events with rotating stellar-mass black holes. In each subfigure,
the light curves (left panels) and astrometric trajectories (right
panels) without and with considering the effect of lens orbital
motion are shown with green dashed and blue dotted lines
respectively. The point-mass lens models are shown by red solid
lines. The photometric and astrometric residuals with respect to the
static binary model are plotted with gray dotted lines. The
parameters of these microlensing events can be found in Table
(\ref{tab2}). We use the generalized version of the adaptive
contouring algorithm \cite{Dominik2007} for plotting astrometric
trajectories in binary microlensing events. Even though, the orbital
motion effect is not obvious in the light curves, this effect can be
detected in the astrometric trajectories. The orbital motion of
lenses makes the astrometric trajectory rotate in the same way as
the source trajectory. 

In the subfigure $(b)$ some oscillatory fashions can be seen in the
astrometric trajectory which is owing to lens orbital motion.
Indeed, whenever the source trajectory does not cross the caustic
curve, the astrometric centroid shift can be estimated as the case
of point-mass lens, i.e. $\delta\theta'_{c}=\frac{u'}{u'^{2}+2}$
where $u'=\sqrt{x'^{2}_{s}+y'^{2}_{s}}$, $x'_{s}$ and $y'_{s}$ are
the components of angular source-lens distance by considering the
effect of lens orbital motion:
\begin{eqnarray}
x'_{s}&=&\sqrt{1-\varepsilon^2} \sin\xi(t) \boldsymbol{[}u_{0}(\cos\beta \cos\alpha-\sin\alpha\sin\beta\sin\gamma)\nonumber\\
&+&p(t)(\cos\beta\sin\alpha+\sin\beta\sin\gamma\cos\alpha)\boldsymbol{]}\nonumber\\
&+&\cos\xi(t) \cos\gamma[p(t)\cos\alpha-u_{0}\sin\alpha]-p(t)\varepsilon \nonumber\\
y'_{s}&=&\sqrt{1-\varepsilon^2} \sin\xi(t) \boldsymbol{[}u_{0}(\cos\beta\sin\alpha+\cos\alpha\sin\beta\sin\gamma)\nonumber\\
&+&p(t)(-\cos\beta\cos\alpha+\sin\beta\sin\gamma\sin\alpha)\boldsymbol{]}\nonumber\\
&+&\cos\xi(t)\cos\gamma[u_{0}\cos\alpha+p(t)\sin\alpha]-u_{0}\varepsilon\cos\gamma,
\end{eqnarray}
where $\alpha$ is the angle between the source trajectory and the
binary axis. By considering the orbital motion some oscillatory
fashions, terms containing $\sin\xi$ and $\cos\xi$, will appear in
the astrometric trajectories which depending on the ratio of the
Einstein crossing time to the orbital period they or some parts of
them can be detected.

When the angle between the orbital motion plane of lenses and sky
plane is about $\sim\pi/2$ i.e. $\gamma\sim0$ and $\beta\sim\pi/2$,
the rotation of binary axis is so small. In that case, the orbital
motion signature in the astrometric trajectories is ignorable
especially when there is no caustic-crossing feature.

\begin{figure}
\begin{center}
\psfig{file=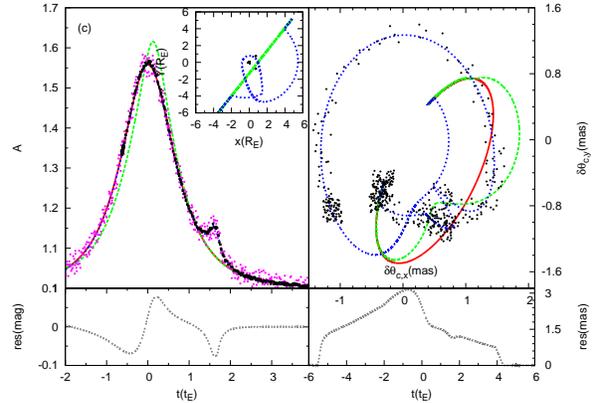,angle=270,width=8.cm,clip=} \caption{A
simulated microlensing light curve and astrometric trajectory of
source star. Data points taken by survey telescopes and the HST are
shown with violet points and black stars. The parameters used to
make this event can be found in table (\ref{tab2}).} \label{fig4}
\end{center}
\end{figure}

Depending on the size of the angular Einstein radius this effect can
be seen in the astrometric trajectory. If there is no
caustic-crossing feature, the probability of detecting the
photometric signature of orbital motion is so small.

\section{Monte Carlo simulation}
\label{Mont} In this section we perform a Monte Carlo simulation to
obtain quantitatively detectability of lens orbital motion in
astrometric and photometric microlensing with rotating stellar-mass
black holes. In the first step, we produce an ensemble of binary
microlensing events with stellar-mass black holes as microlenses
according to the physical distribution of parameters. Then,
corresponding light curves and astrometric trajectories are
generated according to the real data points from an ensemble of
observatories in the microlensing experiment. By considering a
detectability criterion, we investigate whether the signatures of
lens orbital motion can be seen in observations of microlensing
light curves and astrometric trajectories. Our criterion for
detectability of orbital motion is $\Delta
\chi^{2}=\chi^{2}_{OM}-\chi^{2}_{SB}> \Delta\chi^{2}_{th}$ where
$\chi^2_{OM}$ and $\chi^2_{SB}$ are the $\chi^{2}$s of the known
rotating binary microlensing model and the static model with the
same binary parameters respectively. We assume that from fitting
process and searching all parameter space the best-fitted solution
is the known binary solution. Our aim is to compare photometric
efficiency of detecting lens orbital motion with the astrometric
one. Hence, we calculate $\Delta\chi^{2}$ for simulated
light curves and astrometric trajectories independently. 

Here, we explain distribution functions of the lens and source
parameters. For lens parameters we take the mass of the stellar-mass
black hole as the primary from the following distribution function
\cite{Farr2011}:
\begin{eqnarray}
P(M_{l})\propto \exp(-\frac{M_{l}}{M_{o}}),
\end{eqnarray}
where $P(M_{l})=dN/dM_{l}$ and $M_{o}=4.7 M_{\odot}$ in the range of
$M_{l}\in[4.5,25]M_{\odot}$. The mass ratio of the secondary to the
primary is drawn from the distribution function \cite{Duquennoy91}:
\begin{eqnarray}
\rho(q) \propto \exp(-\frac{(q-q_{0})^2}{2\sigma_{q}^{2}})
\label{massratio}
\end{eqnarray}
where $\rho(q)={dN}/{dq}$ in the range of $q\in[0.01,1]$,
$q_{0}=0.23$ and $\sigma_{q}=0.42$. We choose the semi-major axis
$s$ of the binary orbit from the \"{O}pik law where the distribution
function for the primary-secondary distance is proportional to
$\rho(s)=dN/ds\propto s^{-1}$ \cite{Opik} in the range of
$s\in[0.6,30]Au$. The location of lenses from the observer is
calculated from the probability function of microlensing detection
$d\Gamma/dx\propto \rho(x)\sqrt{x(1-x)}$, where $x=D_{l}/D_{s}$
changes in the range of $x\in [0,1]$ and $\rho(x)$ is the stellar
density of thin disk, chosen from Rahal et al. (2009). The time of
arriving at the perihelion point of orbit $t_{p}$ is chosen
uniformly in the range of $t_{p}\in [t_{0}-P,t_{0}+P]$. The
projection angles to indicate the orientation of orbit of lenses
with respect to the sky plane, i.e. $\beta$ and $\gamma$, are taken
uniformly in the range of $[-\frac{\pi}{2},\frac{\pi}{2}]$. We take
the eccentricity of the lenses' orbit uniformly in the range of
$\varepsilon\in[0,0.15]$.

For the source, we take the coordinate toward the Galactic bulge
$(l,b)$, distribution of the matter in standard Galactic model and
generate the distribution of the source stars according to the
Besancon model \cite{dwek95,besancon}. We assign the the absolute,
apparent color and magnitude of the source star in the same approach
which is explained in \cite{sajadian2013}. The mass of source star
is taken from the Kroupa mass function, $\xi = {dN}/{dm} \propto
m_{\star}^{-\alpha}$, in the range of
$m_{\star}\in[0.3,3]M_{\odot}$, where $m_{\star}$ is the mass of the
source star in the unit of the solar mass and $\alpha=0.3$ for
$0.01\leq m_{\star} \leq 0.08$, $\alpha=1.3$ for $0.08\leq
m_{\star}\leq 0.5$ and $\alpha=2.35$ for $m_{\star}\geq 0.5$
\cite{kroupa93,kroupa01}. Another parameter we need in our
simulation is the radius of source star. This parameter will be used
in generating the light curve of microlensing events with the finite
size effect. For the main-sequence stars, the relation between the
mass and radius is given by $R_{\star}= m_{\star}^{0.8}$ where all
parameters normalized to the sun's value. We did not consider giant
source stars. For indicating the trajectory of the source with
respect to the binary lens, we identify this path with an impact
parameter and orientation defined by an angle between the trajectory
of source star and the binary axis. We let this angle change in the
range of $[0,2 \pi]$. The impact parameter is taken uniformly in the
range of $u_0\in[0,1]$ and the corresponding time for $u_0$ is set
zero.

The velocities of the lens and the source star are taken from the
combination of the global and dispersion velocities of the Galactic
disk and bulge \cite{ero09,bin}. The relative velocities of the
source-lens is determined by projecting the velocity of the source
star into the lens plane.

\begin{figure}
\begin{center}
\psfig{file=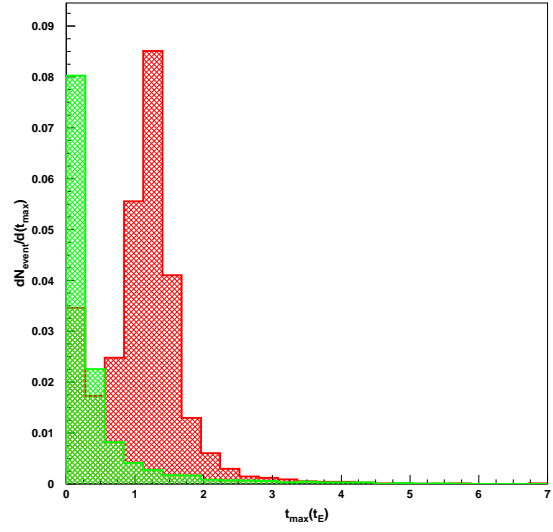,angle=0,width=8.cm,clip=}
\caption{Distribution functions of the time, with respect to the
time of the closest approach and in the unit of the Einstein
crossing time, of the maximum photometric (green histogram) and
astrometric (red histogram) deviations due to lens orbital motion
for binary microlensing events with detectable effects of lens
orbital motion in their light curves and astrometric trajectories
respectively.} \label{fig3}
\end{center}
\end{figure}

We ignore the microlensing events in which the minimum distance
between two lenses and also minimum impact parameter are not one
order of magnitude larger than the Schwarzschild radius of the
primary lens. Because they do not obey Keplerian laws.

Now, we generate data points over the light curves and astrometric
trajectories. We assume that microlensing events are observed with
the HST, OGLE and MOA surveys. The time interval between data points
and the photometric uncertainties for each data point taken by
survey telescopes are drawn from the archive of the microlensing
light curves. The HST start observing these events after that the
light curve reaches to a given threshold of magnification (i.e.
$\frac{3}{\sqrt{5}}$). We suppose that the HST is monitoring these
microlensing events with cadence one day. The other factor in
simulation of the light curve is the exposure time for each data
point taken by the HST. From the exposure time we can calculate the
the error bar for each data point. We can tune the exposure time in
such a way that we have uniform error bars thought out the light
curve. Here, we demand the photometric accuracy for each data point
taken by the HST about $0.4$ per cent which is chosen according to
the recorded exposure times for those long-duration microlensing
events being observed by the HST.
\begin{figure}
\begin{center}
\psfig{file=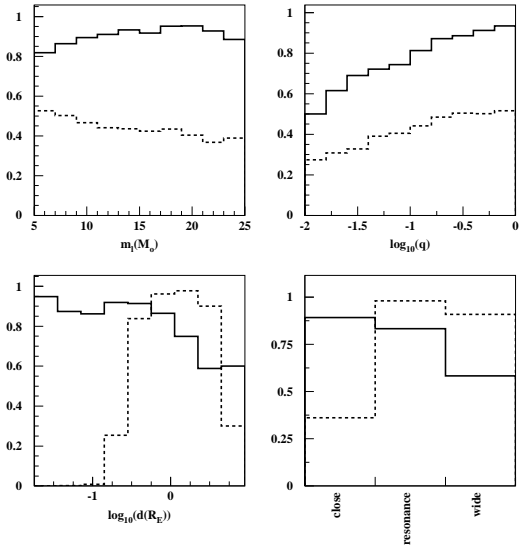,angle=0,width=8.cm,clip=}\caption{The
astrometric (solid line) and photometric (dashed line) efficiencies
of detecting lens orbital motion for different parameters of lenses
and different topology of caustics (the last panel).} \label{fig5}
\end{center}
\end{figure}

Astrometric observations of source star position are done by the HST
with the astrometric accuracy of $\sigma_{as}=200$ micro-arcsecond.
The simulated astrometric data points are shifted according to the
astrometric accuracy of the HST by a Gaussian function. A sample of
the simulated light curve and astrometric trajectory of source star
is shown in Figure (\ref{fig4}). In this figure the data taken by
survey telescopes and the HST are shown with violet points and black
stars. The parameters used to make this event are brought in the
fourth row of table (\ref{tab2}). The time-variation rate of the
astrometric centroid shift vectors is not constant. As a result, in
some places over the astrometric trajectory the HST data points do
not uniformly set despite similar cadences (see Figure \ref{fig4}).
In this event, the orbital motion effect is obvious in the
astrometric trajectory as well as the photometric light curve.

Our criterion for detectability of orbital motion is $\Delta
\chi^{2}=\chi^{2}_{OM}-\chi^{2}_{SB}> \Delta\chi^{2}_{th}$. We
consider $\Delta\chi^{2}_{th}=250$ for both photometric and
astrometric observations.  From the Monte Carlo simulation we obtain
that $80.2$, $16.9$ and $2.9$ per cent of total simulated events are
close, intermediate and wide binaries and the average detection
efficiency $<\epsilon_{OM}>$ for the orbital motion detection in the
astrometric trajectories and the amplification light curves of
binary microlensing events with stellar-mass black holes are $87.3$
and $48.2$ per cent, respectively. As a result, in these events lens
orbital motion can be seen most likely in astrometric trajectories
whereas photometric signatures of orbital motion are often too small
to be detected. Detecting lens orbital motion in their astrometric
trajectories helps to discover further secondary components around
the primary lens as well as resolve close/wide degeneracy.

In Figure (\ref{fig3}) we plot the distribution functions of the
time of the maximum photometric (green histogram) and astrometric
(red histogram) deviations due to lens orbital motion, measured with
respect to the time of the closest approach and in the unit of the
Einstein crossing time i.e. $t_{max}$, for binary microlensing
events with detectable effects of lens orbital motion in their light
curves and astrometric trajectories respectively. According to this
figure, the orbital motion signature in the astrometric trajectory
can be seen at sometime often after one Einstein crossing time with
respect to the time of the closest approach whereas this signal in
the light curve can be detected at sometime almost before it. In
wide or intermediate binary events with detectable astrometric
signatures of orbital motion, the maximum deviation happens at very
late time, e.g. $t-t_{0}\geq 6 t_{E}$, which have made some so small
peaks in Figure (\ref{fig3}).

\begin{figure}
\begin{center}
\psfig{file=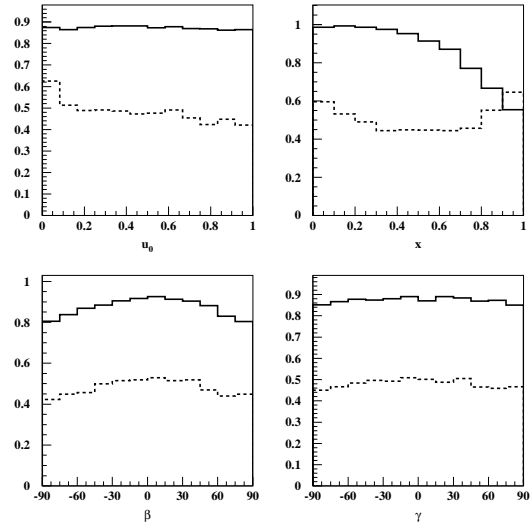,angle=0,width=8.cm,clip=}\caption{The
astrometric (solid line) and photometric (dashed line) efficiencies
of detecting lens orbital motion for different parameters of source
star and lenses trajectories.} \label{fig6}
\end{center}
\end{figure}

In order to study the sensitivity of detecting orbital motion in
astrometric trajectories of source star and photometric light curves
on the parameters of the model, we plot the astrometric (solid line)
and photometric (dashed line) detection efficiencies in terms of the
relevant parameters of the binary lens and source in Figures
(\ref{fig5}) and (\ref{fig6}). We ignore the irrelevant parameters
that do not change the efficiency function. The detection efficiency
function in terms of the binary lens and source parameters is given
as follows.

(i) The first parameter is the primary lens mass, $M_{l}$. Here, the
astrometric efficiency of detecting lens orbital motion rises with
increasing the primary lens mass. The physical interpretation of
this feature is that the angular Einstein radius and as a result of
it the astrometric signal rise with increasing the primary lens
mass. On the photometric detection efficiency there are two factors
that effect inversely. With increasing the primary lens mass the
Einstein radius rises which decreases the normalized distance
between two lens components and the caustic size. On the other hand,
the ratio of the Einstein crossing time to the orbital period of
lenses motion increases. However, the first effect is dominant.
Because, the photometric signature of orbital motion can more likely
be detected while the source is crossing the caustic.

(ii) The second parameter is the ratio of the lens masses, $q$. The
ratio of the lens masses has a geometrical effect on the shape of
the caustic lines, where increasing it towards the symmetric shape
maximizes the detection efficiency. Indeed, the size of the central
caustic scales as $q$ \cite{Gaudi}.

(iii) The third parameter is the projected Semi-major axis of lens
orbit normalized to the Einstein radius. The photometric detection
efficiency has the maximum amount around $1R_{E}$ in which the
caustic curves have the maximum size. However, the astrometric
detection efficiency has the highest amount for close binaries with
the large ratio of the Einstein crossing time to the orbital period.

(iv) Different topologies of caustic curve: The astrometric and
photometric detection efficiencies are maximized in close and
intermediate binaries respectively. Indeed, the size of the central
caustic which indicates the probability of caustic crossing and the
detectability of lens orbital motion photometrically has the highest
amount for the intermediate microlensing events (see Figure
\ref{fig1}). Note that most simulated events are close binaries.

(v) Impact parameter, $u_{0}$, shown in Figure (\ref{fig6}):
Decreasing the impact parameter increases the photometric detection
efficiency. A smaller impact parameter from the center of the lens
configuration rises the probability of the caustic crossing.
However, the astrometric detection efficiency does not depend on the
impact parameter.

(vi) The next parameter is $x=D_{l}/D_{s}$ the relative distance of
the lens to the source star from the observer. The angular Einstein
radius and astrometric signals increase by decreasing $x$. On the
other hand, the Einstein radius has the highest amount around
$x=0.5$. When $x$ tends to $0.5$, by increasing the Einstein radius
the normalized distance between two lens components and the
probability of the caustic crossing decrease.

(vii) and (viii) Inclination angles of orbital plane of lenses with
respect to the sky plane, $\beta$ and $\gamma$. The astrometric
detection efficiency is maximized in $\beta\sim0$ and $\gamma\sim\pm
\pi/2$ in which the rotation angle of the source trajectory with
respect to the binary axis owing to lens orbital motion,
$\theta_{l}$, has the maximum amount.

\section{Conclusions}
\label{result} In this work we investigated lens orbital motion in
astrometric microlensing and its detectability. In microlensing
events, the astrometric centroid shift in source trajectories falls
off much more slowly than the light amplification as the source
distance from the lens position increases. Hence, perturbations
developed with time e.g. the effect of lens orbital motion make
considerable deviations in astrometric trajectories whereas the
photometric signal of orbital motion can mostly be detected when the
source is around the caustic curve.

The orbital motion of lenses changes the magnification pattern by
changing the projected distance between two lenses as well as
rotates the straightforward source star trajectory with respect to
the binary axis. These effects can be seen in the microlensing light
curve if lenses rotate each other considerably when the source star
is around the caustic curve. At the same time, the orbital motion of
lenses rotates the astrometric trajectory in the same was as the
source trajectory. Depending on the size of the angular Einstein
radius this effect can be seen at sometime often after one Einstein
crossing time with respect to the time of the closest approach while
this signal in the light curve can be detected at sometime almost
before it. 

Binary microlensing events with rotating stellar-mass black holes
which have detectable astrometric shifts in the source star
trajectories have most likely detectable orbital motion signatures.
In these events, the ratio of the Einstein crossing time to the
orbital period is high. Although, orbital motion effect in their
light curves is too small especially when there is no
caustic-crossing feature, but the astrometric signature owing to the
orbital motion can be detected at the end parts of astrometric
trajectories. In addition, the Hubble Space Telescope (HST) is
observing some long-duration microlensing events which are likely to
be owing to stellar-mass black holes, following the approved
proposal by Sahu et al. (2010). Hence, we expect that if there is a
secondary lens, the astrometric signature of its orbital motion can
much probably be observed.

By performing Monte Carlo simulation we evaluated the efficiency of
detecting orbital motion in astrometric and photometric microlensing
with binary stellar-mass black holes. We considered $\Delta
\chi^{2}>250$ as the detectability criterion and concluded that
astrometric efficiency is $87.3$ per cent whereas the photometric
efficiency is $48.2$ per cent. From total simulated events $80.2$,
$16.9$ and $2.9$  per cent were close, intermediate and wide
binaries respectively. Also, the more massive lenses, the more
detectable signatures of orbital motion. Detecting orbital motion in
binary microlensing events helps not only to detect further
secondary components even without any photometric binarity signature
but also to resolve close-wide degeneracy as well.

\textbf{Acknowledgment} I am especially thankful to Sohrab Rahvar
for his encouragement, useful discussions and reading of the
manuscript. I would like to thank Martin Dominik for generalizing
his adaptive contouring algorithm to calculate the astrometric
shifts in binary microlensing events, Matthew Penny for pointing out
an important error, Andy Gould and Cheongho Han for careful reading
and commenting on the manuscript. Finally, I thank the referee for
useful comments and suggestions which certainly improved the
manuscript.



\begin{thebibliography}{}

\bibitem[An et al. 2002]{An2002}
An J. H., et al., \ 2002, ApJ,  572, L521.

\bibitem[Binney \& Tremaine 1987]{bin}
Binney, S., \& Tremaine, S. 1987, Galactic Dynamics. Princeton Univ.
Press, Princeton, NJ.

\bibitem[Chang \& Refsdal 1979]{ChangRefesdal}
Chang K., Refsdal S., \ 1979, Nature, 282, L561.

\bibitem[Chung et al. 2009]{Chung} 
Chung S.-J., Park B.-G., Ryu Y.-H. \& Humphrey A. \ 2009, APJ, 695,
L1357.

\bibitem[Dominik 2007]{Dominik2007} 
Dominik M., \ 2007, MNRAS, 377, L1679.

\bibitem[Dominik \& Sahu 2000]{DominikSahu} 
Dominik M. \& Sahu K. C., \ 2000, ApJ., 534, L213.

\bibitem[Dominik 1999]{Dominik99} 
Dominik M. \ 1999, A \& A, 349, L108.

\bibitem[Dominik 1998]{Dominik98} 
Dominik M., \ 1998, A \& A, 329, L361.

\bibitem[Dong et al. 2009]{Dong2009}
Dong S., Gould A., Udalski A., et al., \ 2009, ApJ, 695, L970.

\bibitem[Duquennoy \& Mayor 1991]{Duquennoy91}
Duquennoy A. \& Mayor M., \ 1991, A \& A, 248, L485.

\bibitem[Dwek et al. 1995]{dwek95}
Dwek, E., Arendt, R. G., Hauser, M. G., et al., \ 1995, ApJ, 445,
L716.

\bibitem[Einstein 1936]{Einstein36}
Einstein A., \ 1936, Science, 84, L506.

\bibitem[Farr et al. 2011]{Farr2011}
Farr W. M., Sravan N., Cantrell A., Kreidberg L., Bailyn C. D.,
Mandel I. \& Kalogera V., \ 2011, ApJ, 741, L103.

\bibitem[Gaudi et al. 2008]{Gaudi2008}
Gaudi B. S., et al., \ 2008, Sci, 319, L927.

\bibitem[Gaudi 2012]{Gaudi} 
Gaudi B. S., \ 2012, Annu. Rev. Astron. Astrophys., 50, L411.

\bibitem[Gould \& Han 2000]{Gould}
Gould A. \& Han C. \ 2000, ApJ, 538, L653.



\bibitem[Gould et al. 2013]{Gould13}
Gould A., Shin I. -G., Han C., Udalski A. \& Yee J. C., \ 2013, ApJ,
768, L126.

\bibitem[Han et al. 1999]{Hanetal99}
Han C.,Chun M.-S. \& Chang K. \ 1999, ApJ, 526, L405.


\bibitem[H{\o}g et al. 1995]{Hog}
H{\o}g, E., Novikov, I. D., \& Polnarev, A. G. \ 1995, A \& A , 294,
L287.

\bibitem[Ioka et al. 1999]{Ioka99}
Ioka K., Nishi R. \& Kan-Ya Y., Prog. Theor. Phys., 102, L983.


\bibitem[Jeong et al. 1999]{Jeong}
Jeong Y., Han C. \& Park S.-H. \ 1999, ApJ,511, L569.

\bibitem[Kroupa et al. 1993]{kroupa93}
Kroupa, P., Tout, C. A., \& Gilmore, G.\ 1993, MNRAS, 262, L545.

\bibitem[Kroupa 2001]{kroupa01}
Kroupa, P.\ 2001, MNRAS, 322, L231.

\bibitem[Liebes 1964]{Leibes}
Liebes, Jr.S., \ 1964,  Phys. Rev., 133, L835.

\bibitem[Miralda-Escd\'e 1996]{Miralda96}
Miralda-Escud\'e J., \ 1996, ApJ, 470, L113.

\bibitem[Miyamoto \& Yoshii 1995]{Miyamoto}
Miyamoto, M. \& Yoshii, Y. \ 1995, AJ, 110, L1427.

\bibitem[\"{O}pik 1924]{Opik}
\"{O}pik, E., 1924, Pulications de. L'Observatoire Astronomique de
I'Universit\'{e} de Tartu, 25, L6.

\bibitem[Paczy\'nski 1986a,b]{Paczynski86} 
Paczy\'ski B., \ 1986a, ApJ, 301, L502.

\bibitem[]{Paczynski86b}
Paczy\'ski B., \ 1986b, ApJ, 304, L1.

\bibitem[Paczy\'nski 1995]{Paczynski95} 
Paczy\'ski B., \ 1995, Acta Astron., 45, L345.

\bibitem[Paczy\'nski 1997]{Paczynski97}
Paczy\'nski B., \ 1997, Astrophys. J. Lett. astro-ph/9708155.

\bibitem[Penny et al. 2011]{penny11}
Penny M. T., Mao S. \& Kerins E., \ 2011, MNRAS, 412, L607.

\bibitem[Proft et al. 2011]{GAIA}
Proft A., Demleitner M. \& Wambsganss J., \ 2011, A \& A, 536, L50.

\bibitem[Rahal et al. 2009]{ero09}
Rahal, Y. R., Afonso C., Albert J.-N., et al. \ 2009, A \& A, 500,
L1027.

\bibitem[Rattenbury et al 2002]{Rattenbury}
Rattenbury N. J., Bond I. A., Skuljan J. \& Yock P. C. M., \ 2002,
MNRAS, 335, L159.

\bibitem[Robin et al. 2003]{besancon}
Robin, A. C., Reyl\'{e}, C., Derri\`{e}re, S., Picaud, S., \ 2003, A
\& A, 409, L523.

\bibitem[Sajadian et al. 2013]{sajadian2013}
Sajadian, S., Rahvar, S. \& Dominik M., \ 2013, in preperation.

\bibitem[Sahu et al. 2010]{Sahuproposal} 
Sahu K. C., Bond H. E., Anderson J., Udalski A., Dominik M. \& Yock
P., \ 2010, the proposal number: 13458.

\bibitem[Sahu 2011a]{Sahu} 
Sahu K. C., \ 2011, Oral Presentation at 15th Microlensing Workshop,
Salerno.

\bibitem[Sahu 2011b]{Sahu2}
Sahu K. C., \ 2011, Oral Presentation at annual MindStep meeting,
Qatar.

\bibitem[Shin et al. 2011]{Shin11}
Shin I. -G., Udalski A., Han C. et al. \ 2011, ApJ, 735, L85.

\bibitem[Shin et al. 2012]{Shin12}
Shin I. -G., Han C., Choi J. -Y., et al. \ 2012, ApJ, 755, L91.

\bibitem[Shin et al. 2013]{Shin2013} 
Shin I. -G., Sumi T., Udalski A., et al., \ 2013, ApJ, 764, L64.



\bibitem[Walker 1995]{Walker} 
Walker M. A. \ 1995, ApJ, 453, L37.

\end {thebibliography}
\end{document}